# Reconstructing phonon mean free path contributions to thermal conductivity using nanoscale membranes

*John Cuffe[1], Jeffery K. Eliason[2], Alexei A. Maznev[2], Kimberlee C. Collins[1], Jeremy A. Johnson[1†],*

*Andrey Shchepetov[3], Mika Prunnila[3], Jouni Ahopelto[3], Clivia M. Sotomayor Torres[4,5],*

*Gang Chen[*1], Keith A. Nelson[2]*

[1]Department of Mechanical Engineering, Massachusetts Institute of Technology,

77 Massachusetts Avenue, Cambridge, MA 02139, USA.

[2]Department of Chemistry, Massachusetts Institute of Technology, 77 Massachusetts Avenue,

Cambridge, MA 02139, USA.

[3]VTT Technical Research Centre of Finland, PO Box 1000, 02044 VTT, Espoo, Finland.

[4]ICN2 - Catalan Institute of Nanoscience and Nanotechnology, Campus UAB, Edifici ICN2,

08193 Bellaterra, Spain

[5]Catalan Institute for Research and Advanced Studies ICREA, 08010 Barcelona, Spain.

ABSTRACT: Knowledge of the mean free path distribution of heat-carrying phonons is key to understanding phonon-mediated thermal transport. We demonstrate that thermal conductivity measurements of thin membranes spanning a wide thickness range can be used to characterize how bulk thermal conductivity is distributed over phonon mean free paths. A non-contact transient thermal grating technique was used to measure the thermal conductivity of suspended Si membranes ranging from 15 to 1500 nm in thickness. A decrease in the thermal conductivity from 74% to 13% of the bulk value is observed over this thickness range, which is attributed to diffuse phonon boundary scattering. Due to the well-defined relation between the membrane thickness and phonon mean free path suppression, combined with the range and accuracy of the measurements, we can reconstruct the bulk thermal conductivity accumulation vs. phonon mean free path, and compare with theoretical models.

Phonon mean free path (MFP) plays a key role in lattice thermal conductivity. Recent studies have demonstrated that describing phonon-mediated heat transport with a single or average MFP value is largely inadequate [1–4]. A more adequate description is provided by considering a distribution of thermal conductivity over phonon mean free paths (MFPs), also known as the phonon MFP distribution or spectrum [5,6]. Knowledge of this distribution is necessary for modeling heat transport in nanostructures and designing new materials for applications such as thermoelectric energy conversion [7]. However, accurate measurements of MFP distributions are challenging, as the length scales of thermal phonon wavelengths and MFPs can vary from less than 1 nm to several micrometers, depending on the material and temperature [8,9]. Experimental measurements of thermal conductivity yield only an integrated value, and do not typically reveal microscopic details of the thermal transport. First-principle calculations of thermal conductivity accumulation vs MFP have recently been accomplished for a few materials [8–10]. However, the complexity of modeling anharmonic phonon-phonon interactions necessitates experimental validation of these calculations.

Although the frequency-dependence of phonon lifetimes can be measured directly with inelastic neutron [11], X-ray [12] or light scattering [13], or with laser-excited coherent phonons [14,15], MFP measurements by these techniques have been limited. For example, the available experimental data for silicon do not extend beyond 100 GHz [16], far below the range of frequencies thought to be important for thermal transport at room temperature [8,10].

In recent years, advances in measuring phonon MFP distributions have been made with studies of size-dependent thermal conductivity over small distances, where thermal transport deviates from Fourier's law. In these experiments, the length scale is typically controlled through the measurement geometry, varying, for example, metal line width [1], laser spot size [2], optical grating period [3], or laser modulation frequency [4]. As the heat flux of phonons with MFPs longer than the measurement length scale is suppressed compared to that predicted by Fourier's heat diffusion model, the observed reduction in effective thermal conductivity can be used to estimate the contributions of phonon MFPs to the heat flux [2–4]. These experiments have provided useful insights into phonon MFP distributions, although there are currently lower limits to the length scales that can be probed, e.g. ~300 nm for crystalline silicon [4]. Moreover, in order to find the MFP distribution from size-dependent thermal transport measurements (the inverse problem), one should be able to calculate thermal transport from a known MFP distribution (the direct problem). For most experimental geometries, the solution to the direct problem has not been obtained; instead, approximations are typically made, such as cutting off the contribution of phonons whose MFP exceeds a certain length. In addition, the interpretation of these non-diffusive effects is often complicated by the presence of a metal-substrate interface.

The most progress in solving the direct problem has been achieved with the laser-induced transient thermal grating (TTG) technique which provides the advantage of a very simple geometry [3]. However, while a sound theory has been developed for a one-dimensional thermal grating in a bulk material [17], the effects of interplay between boundary scattering and non-diffusive MFP suppression in experimental measurements has not been accounted for in a rigorous manner [3].

In this work, we demonstrate an alternative approach to characterizing the MFP distribution based on size-dependent thermal conductivity measurements. Namely, we measure thermal transport in suspended nanoscale membranes in the *diffusive* regime and reconstruct the thermal conductivity accumulation vs MFP from the dependence of the thermal conductivity on the membrane thickness. The advantages of this approach are the availability of the direct solution to the size-dependent heat flux problem based on the Fuchs-Sondheimer model [18–20] and the access to a length scale down to ~10 nm. Studies of nanostructures have shown that thermal conductivity reduces significantly with decreasing thickness due to the reduction of phonon MFPs caused by diffuse boundary scattering [21–23]. While thin-film thermal conductivity has been investigated previously, accurate measurements covering a wide thickness range are needed in order to recover the MFP distribution from size-dependent thermal conductivity data. To further reduce uncertainty, these measurements should, preferably, avoid heat transport across interfaces, and only measure in-plane transport. To meet these requirements, we use the non-contact laser-based TTG technique [3,24], which inherently yields high absolute accuracy, to measure suspended single-crystalline silicon membranes with thickness values ranging from 15 nm to 1.5 μm. Then, following the work of Minnich [5], and Yang and Dames [6], we use a convex optimization algorithm to reconstruct the thermal conductivity accumulation as a function of MFP for bulk silicon at room temperature.

The membranes with areas of ~500 x 500 μm$^2$ were fabricated on 150 mm silicon-on-insulator (SOI) wafers using Si MEMS processing techniques [25]. The underlying Si substrate and the buried oxide layer were removed through a combination of dry and wet etching techniques to leave a top layer of suspended silicon. The high etch selectivity of the buried oxide with respect to the top SOI layer enables the release of the membrane. The thickness was obtained from optical reflectance measurements performed with a FilmTek 2000 spectroscopic reflectometer. The accuracy of the measurements is estimated to be better than one nanometer.

The optical setup used to create a thermal grating in the sample and monitor its decay (Fig. 1a) was similar to that used in prior works [3,24]. A 515 nm pulsed excitation laser with a 60 ps pulse duration at a 1 kHz repetition rate was passed through a transmissive diffraction grating. A two-lens imaging system was then used to collect the ±1 orders of diffraction and cross them at the sample to generate an interference pattern. The period of the interference pattern is determined by the diffraction grating period and the imaging system. Absorption of the pulses results in a spatially periodic temperature variation, which behaves as an optical diffraction grating due to the temperature dependence of the complex refractive index and the membrane thickness. A continuous wave probe laser with a wavelength of 532 nm was passed through the same diffraction grating and imaging system as the pump to generate probe and reference beams. The probe laser was modulated by an electro-optic modulator synchronized to the pump laser repetition rate with a temporal window of 64 μs duration in order to reduce sample heating by the probe light. The probe beam is diffracted from the transient thermal grating and superimposed with the attenuated reference beam for heterodyne detection in order to enhance sensitivity [24]. The signal was monitored with a 1 GHz amplified photodiode and recorded by a digitizing oscilloscope.

As a result of the one-dimensional sinusoidal heating pattern, the temperature grating decays exponentially due to thermal transport from the heated to the unheated regions with a decay time $\tau = \alpha q^2$, where $\alpha$ is the thermal diffusivity and $q = 2\pi/L$ is the grating wave vector magnitude corresponding to the grating period $L$. The thermal diffusivity can therefore be determined from the signal decay time [26],

$$\alpha = 1/q^2\tau. \qquad (1).$$

Both the signal decay time and the grating period can be measured with high accuracy, with no other parameters involved in the measurement; in particular, precise knowledge of the absorbed laser power and the magnitude of the temperature variation is not required, thus eliminating a major challenge to measuring thermal conductivity in nanoscale objects [27,28]. As no adsorbed metal heater layer is used and the samples are suspended, the measurements and analysis are free from uncertainties due to thermal interface resistance and heat loss to a substrate. In recent work, we have observed non-diffusive transport with TTG periods $L$ that are short compared to the MFPs of heat-carrying phonons [3]; in the present work we use large periods so that the film thickness is the only dimension that is made small relative to phonon MFPs and the resulting changes in diffusivity can be associated uniquely to the thickness variations. The absence of non-diffusive effects can be verified by varying the TTG period and ensuring that the measured thermal diffusivity remains constant, i.e., that the transport time scales as the square of the transport length, $\tau \propto 1/q^2 \propto L^2$, as expected for diffusion.

Data were collected for grating periods ranging from 11 to 21 μm for silicon membranes with thickness values ranging from 15 nm to 1.5 μm. The diffusivity was calculated from a biexponential fit to the acquired traces to account for the electronic response at short times [3] followed by the significantly slower thermal transport. The measurements were performed at 294 K in a cryostat under vacuum, thus removing any potential contribution from thermal transport through air. The optical penetration depth of 515 nm light in silicon is approximately 1.2 μm, and the thermal grating periods were much larger than the thickness. This ensured that the thermal grating was nearly homogeneous throughout the thickness of the samples and so thermal transport could be considered to be one-dimensional. The effect of sample heating by the pump and probe lasers was estimated by repeating the experiments with twice the original laser powers. It was found that pump power had a negligible effect, while the higher probe power resulted in a reduction of the measured diffusivity typically less that 5%. The uncertainties associated with the probe power are included in the error bars in Fig. 2b as positive uncertainties, as the increased power reduced the measured diffusivity.

Figure 1b shows typical signal intensities as functions of time for silicon membranes with thickness values of 1.5 μm, 100 nm and 17.5 nm for a thermal grating period of 21 μm. The thermal diffusivity is determined from the decay time according to Eq. (1) and the thermal conductivity can then be calculated as $k = \alpha C$ where $C = 1.64 \times 10^6$ J m$^{-3}$ K$^{-1}$ is the volumetric heat capacity of silicon [29]. The volumetric heat capacity of the membranes is predicted not to change significantly for silicon membranes with thickness values down to 15 nm at room temperature, due to the relatively small change in the density of states [30,31].

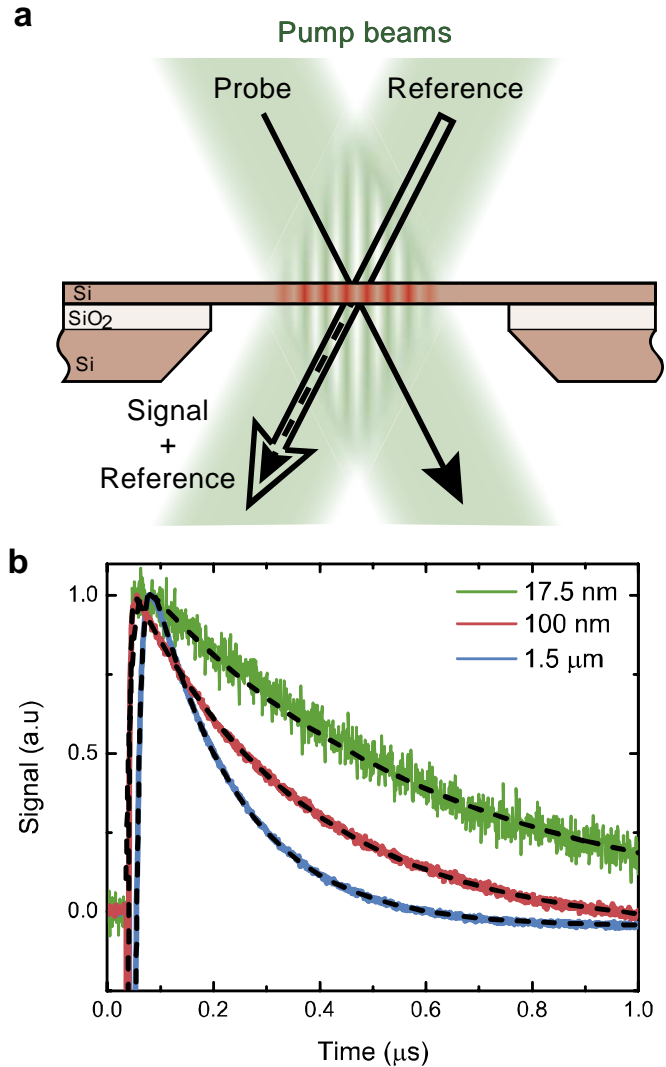

FIG. 1. (a) Schematic illustration of the Si membrane samples and TTG experiment. Two laser pump pulses are crossed on the suspended membrane to form a thermal grating. A probe beam is diffracted from the thermal grating and mixed with a reference beam for heterodyne detection. (b) Typical signals as functions of time for silicon membranes with thickness values of 1.5 μm (blue), 100 nm (red) and 17.5 nm (green) at a grating spacing of 21 μm, showing a slower decay for the thinner membranes. The fitted thermal decay (dashed line) is related to the thermal diffusivity of the sample as described in the text.

Figure 2a shows that the measured diffusivities of the membranes remain constant as a function of the grating spacing, which is indicative of diffusive thermal transport. Figure 2b shows the associated thermal conductivities as a function of thickness, compared to other experimental works [32–34]. The plotted thermal conductivity for each membrane thickness is the average value calculated from all grating spacings. The error bars include statistical variation between grating spacings and the estimated effect of sample heating.

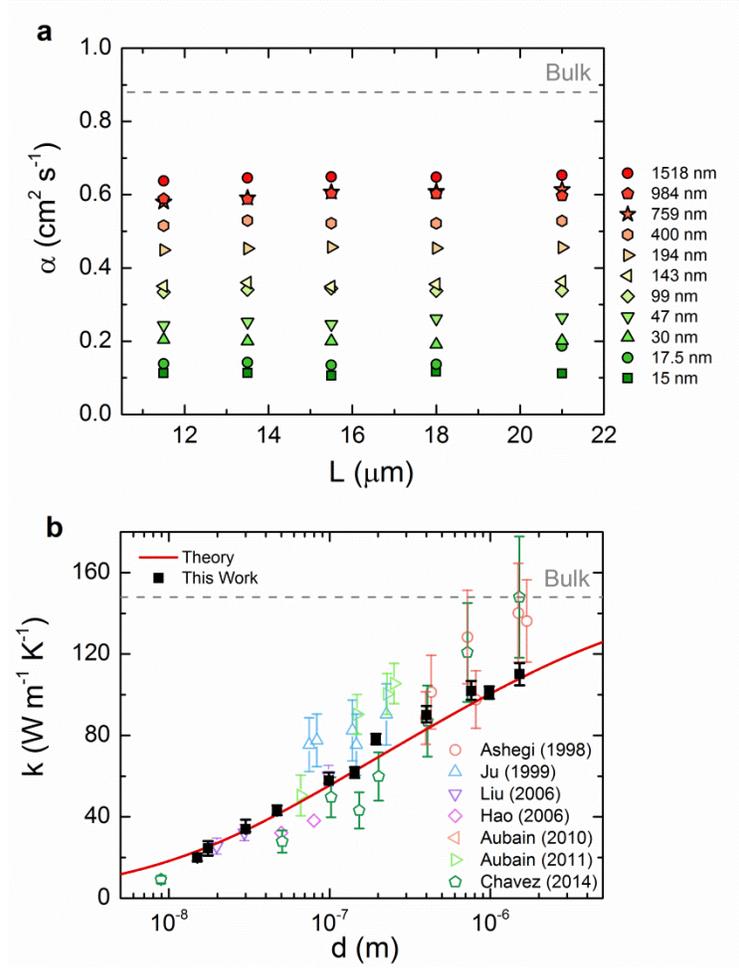

FIG. 2. Thermal conductivity characterization of the Si membranes. (a) Thermal diffusivities $\alpha$ as functions of grating spacing $L$ for silicon membranes with thickness values ranging from 15 nm to 1.5 μm. (b) Thermal conductivity $k$ as a function of membrane thickness $d$. The experimental points from this work (black squares) are in good agreement with the calculation based on Eq. (4) with the phonon MFP distribution calculated from first principles [8] (red line). Data from other thermal conductivity measurements on supported and unsupported Si thin films are shown for comparison in open symbols [28,33–38]. The thermal diffusivity and conductivity values of bulk silicon are shown for reference in (a) and (b) respectively [29,39].

To compare our experiments with theory, we calculate the thermal conductivity in the membrane using the phonon Boltzmann transport equation, assuming isotropic dispersion under the relaxation time approximation [40],

$$k_{mem} = \sum_s \int_0^\infty \frac{1}{3} C_\omega \, v \, S\left(\frac{\Lambda_{bulk}}{d}\right) \Lambda_{bulk} d\omega. \tag{2}$$

This equation describes the thermal conductivity in terms of the individual contributions from phonons with frequency ω, with a volumetric heat capacity $C_\omega$, group velocity $v$ and a MFP Λ, summed over each polarization branch *s*. The function S(Λ$_{bulk}$/d) represents how the contribution to the heat flux of a phonon with a MFP *Λ$_{bulk}$* is suppressed compared to the bulk for a membrane of thickness *d* and is given by

$$S\left(\frac{\Lambda_{bulk}}{d}\right) = 1 - \frac{3}{8}\frac{\Lambda_{bulk}}{d} + \frac{3}{2}\frac{\Lambda_{bulk}}{d}\int_1^\infty \left(\frac{1}{t^3} - \frac{1}{t^5}\right) e^{-\frac{d}{\Lambda_{bulk}} t} dt, \tag{3}$$

This relation, known as the Fuchs-Sondheimer [19] suppression function, was originally derived analytically from the Boltzmann transport equation to calculate thin film electrical conductivity. Although this equation is commonly used to describe the lattice thermal conductivity of thin films [23], its full derivation for phonon-mediated heat transport with associated assumptions is not found in the literature. We provide the derivation the supplementary information [20]. A similar expression has been derived by Turney et al. [41].

In deriving Eq. (2), we have assumed that the discretization of the z-component of the wave vector due to the finite film thickness is negligible, or, in other words, that there are no phonon confinement effects apart from the reduction of the mean free path due to boundary scattering, which we have assumed to be completely diffuse in deriving Eq. (3). An accurate description of wave scattering from rough surfaces is quite complex [42], and requires knowledge of the surface roughness and a model for describing the wavelength dependence of the specularity, p. If we assume that the majority contribution to the thermal conductivity of Si at room temperature comes from phonons with wavelengths of < 6 nm [8], and extrapolate recent experimental results from photo-acoustic measurements of the specularity of Si membranes at ultrasonic phonon wavelengths we find that p = 0 is a reasonable assumption [42].

When the contributions of long wavelength phonons are important, as expected, for example, at low temperatures, accounting for the wavelength-dependent specularity may be required [42,43]. Additionally, when phonon wavelengths are comparable to the film thickness, a modification of the dispersion relation occurs. While it is challenging to quantitatively predict the effect of this modification on thermal conductivity, we can estimate the effect by comparing the lattice thermal energy, $E_{TH} = k_B T$, and the energy spacing between the phonon branches, $\Delta E = \hbar \Delta \omega$. At the centre of the Brillouin zone this separation is approximately $\Delta E = \hbar \pi v_i / d$ where $v_i$ can be the longitudinal or transverse sound velocity. For high temperatures (> 15 K) and thick membranes (> 15 nm), the energy between the modes is always much smaller than the thermal lattice energy, and so many modes are occupied [44]. In this case the dispersion relation and the phonon density of states can be well approximated by their bulk values.

We perform a change of variables in Eq. (2) to express the thermal conductivity in the membrane as:

$$k_{mem} = \int_0^\infty K_{\Lambda_{bulk}} S\left(\frac{\Lambda_{bulk}}{d}\right) d\Lambda_{bulk} , \qquad (4),$$

where $K_{\Lambda_{Bulk}} = -\sum_s \frac{1}{3} C v \Lambda_{bulk} \frac{d\omega}{d\Lambda_{bulk}}$ is the thermal conductivity contribution per MFP, known as the differential MFP distribution [6]. If the MFP distribution is known, the thermal conductivity in the membranes can be calculated from Eq. (4) directly, without explicit knowledge of the frequency-dependence of the phonon group velocities and MFPs. For silicon, the MFP distribution calculated from first principles [8] predicts membrane thermal conductivity in good agreement with our experimental data, as shown in Fig. 2b. The importance of this result is that it demonstrates the ability of first-principle-based calculations to predict thermal conductivity of nanostructures without free parameters. Previously, first-principles calculations of lattice thermal conductivity were tested by comparing them to the temperature dependence of the thermal conductivity of bulk single crystal materials [8,45]. However, the true promise of the ab-initio approach is in predicting thermal transport in engineered materials and structures. Another important observation is the fact that the Fuchs-Sondheimer model works well for the thermal conductivity of thin films down to 15 nm in thickness at room temperature. The thickness at which phonon confinement effects become significant at room temperature remains an open question.

Now let us consider the inverse problem of reconstructing the phonon MFP distribution from the experimental measurements. To do this, we rewrite Eq. (4) in terms of the normalized accumulative MFP distribution, defined as:

$$k_{acc}(\Lambda_c) = \frac{1}{k_{bulk}} \int_0^{\Lambda_c} K_{\Lambda_{bulk}} \, d\Lambda_{bulk} \qquad (5).$$

which represents the fraction of thermal conductivity contributed by all phonons with MFP less than $\Lambda_c$ [5,6]:

$$\frac{k_{mem}}{k_{bulk}} = \int_0^\infty k_{acc}(\Lambda_{bulk}) \frac{dS\left(\frac{\Lambda_{bulk}}{d}\right)}{d\Lambda_{bulk}} \, d\Lambda_{bulk} \qquad (6).$$

While inverting this equation to recover $k_{acc}(\Lambda_c)$ from experimental measurements of $k_{mem}$ as a function of *d* is technically an ill-posed problem, certain constraints can be imposed on the form of the accumulative distribution to allow it to be reconstructed, given a wide enough range of experimental data. Minnich [5] proposed an algorithm based on a convex optimization procedure [46] to find the "smoothest" accumulation function still satisfying Eq. (5) within experimental uncertainties, under the restriction that the function increases monotonically from 0 to 1.

The result of the reconstruction obtained with this algorithm is shown in Fig. 3. It can be seen that the reconstructed distribution is broad and agrees quite well with first principles calculations. We also compare our results with the commonly-used Holland model [47]. While Holland's model predicts a dominant contribution by phonons with MFPs close to 300 nm, the reconstructed distribution is much broader. For example, it can be seen that phonons with MFPs lager than 1 μm contribute almost 50% to the overall thermal conductivity. We note that although the Holland model correctly reproduces the temperature-dependent thermal conductivity of bulk silicon [47], it fails to predict the contributions of phonons with different MFPs, which emphasizes the need for size-dependent thermal conductivity measurements to characterize MFP distributions.

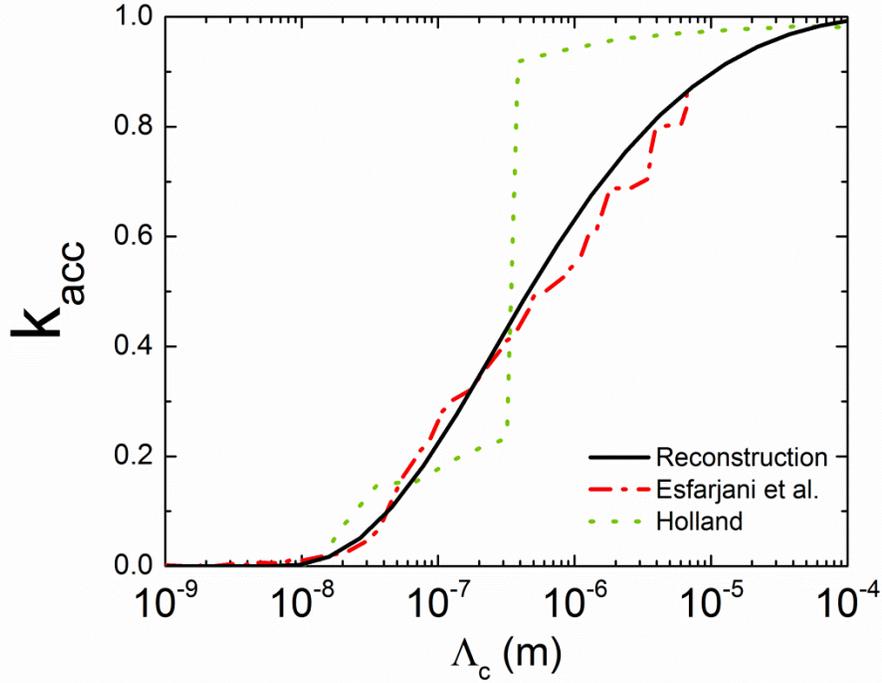

FIG. 3. Reconstructed phonon MFP distribution. Normalized thermal conductivity accumulation $k_{acc}$ showing the fraction of thermal conductivity contributed by phonons with MFPs less than $\Lambda_c$. The reconstructed thermal conductivity accumulation function shows good agreement with the distribution calculated from first principles [8].

Our results show compelling experimental evidence of the broad phonon MFP distribution in silicon, confirming recent *ab initio* calculations. The reconstruction was possible due to the well-defined relation between characteristic dimension of membrane thickness and phonon MFP reduction described by the Fuchs-Sondheimer suppression function, as well as the accuracy and range of the measurements, spanning from 1.5 μm down to 15 nm in thickness. A natural next step will be an investigation of the temperature dependence of the phonon MFP distribution. However, the interpretation of the low temperature data requires a more careful treatment of phonon confinement effects, as well as a wavelength-dependent model of the surface specularity.

Our methodology can be extended to other materials, for which accurate ab-initio calculations may not be available. The capability of characterizing the MFP distribution of heat-carrying phonons will further advance the quantitative understanding of phonon-mediated thermal transport for important fields such as thermoelectrics and nanoelectronics.

AUTHOR INFORMATION

**Corresponding Author**

*gchen2@mit.edu

Present Addresses: †Paul Scherrer Institut, Villigen, Switzerland.

ACKNOWLEDGMENTS

The authors gratefully acknowledge helpful discussions with Emigdio Chavez. We acknowledge support from "Solid State Solar-Thermal Energy Conversion Centre (S3TEC)," an Energy Frontier Research Centre funded by the U.S. Department of Energy, Office of Science, Office of Basic Energy Sciences under Grant No. DE-SC0001299/DE-FG02-09ER46577, the Academy of Finland under grant No. 252598 and the EU FP7 ENERGY FET project MERGING grant agreement No. 309150 and the Spanish Plan Nacional project TAPHOR (MAT-2012-31392).